\documentclass{article}
\usepackage{amssymb}
\usepackage{amsmath}
\usepackage{latexsym}
\usepackage{bm}
\usepackage{graphicx}
\usepackage{lscape}
\usepackage{booktabs}
\usepackage{indentfirst}
\begin{document}

\title{Scaling properties of multiplicity fluctuations in heavy-ion collisions simulated by AMPT model
}
\author{Xie Yi-Long, Chen Gang\footnote{Email:chengang1@cug.edu.cn},Wang jiang-ling, Liu Zhao-Hui, Wang Mei-Juan}
\date{}
\maketitle
\begin{center}
\mbox{Physics Department, School of Mathematics and Physics, China
University of Geosciences} \mbox{Wuhan, China 430074}

\end{center}

\begin{abstract}
Three dimensional, as well as one- and two-dimensional, studies of
multiplicity fluctuation are performed using AMPT model to generate
central Au-Au collision events at $\sqrt{s_{\rm{NN}}}=200$~GeV. Two-
and three-dimensional normalized factorial moments in rapidity,
transverse momentum and azimuthal angle are found to exhibit
power-low scaling when partitioning with the same number of bins in
each direction, indicating that the fluctuation are isotropic, i.e
the fractality is self-similar in multiparticle production of
central Au-Au collisions. Further, we measured the parameter $\nu$
which it characterizes the intermittency indices derived in
particular analysis. It is found that our model result
$\nu_{y\varphi p_t}= 1.86\pm 0.07$ is larger than $\nu=1.304$, which
is the value of Ginzburg-Landau type of phase transition. We also
explored the intermittent or fluctuational dependence on the
transverse momentum. The result shows that the intermittency or
fluctuation increase rapidly with the increasing of transverse
momentum.

\end{abstract}


\section{Introduction}

As has been known, large non-linear fluctuations exist in the
process of space-time evolution in high-energy collisions. Such
large local fluctuations have been observed in a cosmic-ray
event~\cite{JACEE} and in high-energy collision experiments
~\cite{NA221,UA5,bialas1}. Theoretically, the QCD branching
cascade~\cite{QCD1} involving channels $q \rightarrow qg$,
$g\rightarrow gg$ and $g \rightarrow q\bar{q}$, like other branching
processes~\cite{QCD2}, leads to fractal behavior~\cite{QCD3} which
manifests itself in the form of power law scaling of final-state
multiplicity fluctuations with an increasing resolution in phase
space.

In order to be able to decide whether these fluctuations are
dynamical, i.e. larger than expected from Poisson noises, Bialas and
Peschanski~\cite{bialas1} have suggested the use of normalized
factorial moments(NFM), which is:
\begin{eqnarray}
F_q(\delta y)=
\frac{1}{M}\sum_{m=1}^{M}\frac{<n_m(n_m-1)\cdots(n_m-q+1)>}{<n_m>^q}
\end{eqnarray}
where $\delta y=\frac{\Omega}{M}$ is the size of each phase-space
when dividing the whole phase space zone $\Omega$ into $M$ parts,
and $n_{m}$ is the multiplicity in sub-phase space
$\Omega_{m}(\delta y)$. In this way, the dynamic fluctuation in
high-energy collisions can be manifested as an abnormal scaling
property of NFM, i.e.:
\begin{eqnarray}
F_q(\delta y)\propto (\delta y)^{-\phi_q},\quad (\delta y\to 0),
\end{eqnarray}
when the corresponding collision system is a fractral. In general,
for one-dimensional variables, the factorial moments tend to
saturate at small phase-space intervals. This can be explained as a
projection effect of a three-dimensional phenomenon~\cite{Ochs}. It
is therefore expected that the scaling phenomenon can be observed in
a higher-dimensional analysis.

In three-dimensional phase space the anomalous scaling, or fractal,
may be either isotropic or anisotropic, depending on the way a phase
space is partitioned~\cite{Wulls}. If the scaling is observed when
the phase space is partitioned as $\lambda _a=\lambda _b=\lambda
_c$, then the anomalous scaling is isotropic (corresponding to a
self-similar fractal), where $\lambda _a, \ \lambda _b, \ \lambda
_c$ denote shrinking ratios of the three phase-space directions;
otherwise it is anisotropic (self-affine fractal~\cite{QCD2}).

In the 1990s the anomalous scaling of NFM in high-energy collisions
were studied extensively~\cite{DDKreview,WLreview}. In hadron-hadron
collisions, the anomalous scaling of NFM was
observed~\cite{NA22SA,NA27SA} to be anisotropic (self-affine
fractal). However, the anomalous scaling of NFM in e$^+$e$^-$
collisions closely obeyed the scaling properties in Eq. (2) for
isotropic partition of three-dimensional phase
space~\cite{chenli,LLL,DELPHI, Chen}. In our work we applied a
self-similar analysis to heavy ion collisions using the Multi-phase
Transport (AMPT) model ~\cite{linko} to generate central Au-Au
collision events at $\sqrt {s_{\rm{NN}}}=200$~GeV.

In this paper, the AMPT model~\cite{linko} is introduced in Section
2. The method of self-similar or self-affine analysis is briefly
summarized in Section 3. The results of the self-similar analysis
for central Au-Au collisions are shown in Section 4. The FM's
scaling property and the Ginzburg-Landau type of phase transition
applies are discussed in Section 5. Conclusions are presented in
Section 6.

\section{A brief introduction of AMPT}

The AMPT model~\cite{linko} is a mixed model based on both hadronic
and partonic phase and exists in two versions: the default AMPT and
the AMPT with string melt. Each version contains four subprocesses:
phase-space initialization, parton scattering, hadronization and
hadron rescattering. The initialization takes the HIJING model as
the event generator, which includes minijet production and soft
string excitations. Scattering among the partons follows Zhang's
Parton Cascade model, which includes only a simple two-body
scattering. The cross-sections of the partons are calculated by
pQCD. In the default AMPT, the transition from partons to hadrons,
i.e. hadronization, follows the Lund string fragmentation model. In
this situation, when the partons stop interacting with each other,
they melt with their parent strings and are then converted into
hadrons. Conversely, in AMPT with string melting, the minijet
partons melt with their parent stings to become excited strings,
which then fragment into hadrons. Due to the assumption of high
initial energy density in the model, these hadrons melt into valence
quarks and antiquarks. After the ZPC parton cascade, the
hadronization adopts the Quark Coalescence model, in which the two
nearest partons combine to become a meson, and three nearest partons
into a baryon. Finally, the rescattering and resonance decay of the
partons are described by the ART hadronic transport model.

It is well known that the AMPT model with string melting offers a
better description of elliptic flow and $\pi$ correlation function,
while the default AMPT model provides a better simulation of
rapidity distribution and transverse momentum spectrum. In our work,
the phase-space variables such as rapidity, transverse momentum and
azimuth were used to study the fractal characteristics in central
Au-Au collisions. We then utilized the default AMPT to generate
central Au-Au collision events at $\sqrt {s_{\rm{NN}}}=200$~GeV. The
impact parameter is in the range b $\leqslant$ 2 fm and the parton
cross-section is taken to be 3 mb.

\section{The method}
In three-dimensional phase space, it can be determined whether the
anomalous scaling, or fractal, is isotropic or anisotropic depending
on the mode of the phase-space partition. If the scaling, Eq. (2),
is observed when the phase space is partitioned according to:
\begin{equation}
 \Delta x_i\to \delta x_i = \frac{\Delta x_i}{\lambda_i},(i=a,b,c)
 \end{equation}
where $i$ denotes the three phase space directions, then anomalous
scaling is isotropic (corresponding to self-similar fractal) in the
case of $\lambda_a=\lambda_b=\lambda_c$; otherwise it is anisotropic
(self-affine fractal~\cite{Wulls}). Note that $\lambda_i$ is the
phase space representing the partition number or the shrinking
ratios in direction $i$. The three-dimensional partition number is
the product of the 3 $\lambda_i$'s:
\begin{equation}M=\lambda _a \cdot \lambda_b \cdot \lambda_c.
\end{equation}
The shrinking ratios $\lambda_a$, $\lambda_b$ and $\lambda_c$ are
characterized by a parameter
\begin{equation}H_{ij}=\frac{\ln{\lambda_i}}{\ln{\lambda_j}} \quad
(i,j=a,b,c),
\end{equation}
which is called the Hurst exponent~\cite{Hurst}. It is characterized
as the isotropic property of a fractal. If all the three Hurst
exponents are equal to unity, i.e. $H_{ab} = H_{bc} = H_{ca} = 1$,
then the fractal is self-similar; otherwise it is self-affine.

The Hurst exponents can be deduced by fitting three one-dimensional,
second-order NFM saturation curves~\cite{Ochs}:
\begin{equation}F_2^i(\lambda_i)=A_i-B_i\lambda_i^{-\gamma_i}\quad
(i=a,b,c),
\end{equation}
where $\gamma_i$ describes the saturation rate of the NFM along the
direction of $i$. The Hurst exponents are related to the
$\gamma_i$'s as:
\begin{equation}
H_{ij}=\frac{1+\gamma_j}{1+\gamma_i}\quad (i,j=a,b,c).
\end{equation}

Therefore, from the final-state multiplicity production of
high-energy collisions, we can calculate the Hurst exponents that
describe the fractal property of the system, and hence the phase
space can be separated properly~\cite{wu,wu1}.

However, even in the central region the rapidity distribution is not
flat. The particle distribution of final state has a trivial effect
on the scaling behavior of the NFM. Therefore, the cumulant variable
\begin{equation}
X_c=\frac{\int_{X_{min}}^{X}\rho(X)\mathrm{d}X}{\int_{X_{min}}^{X_{max}}\rho(X)\mathrm{d}X}
\end{equation}
was introduced~\cite{li} to reduce the effect of trivial
fluctuations, where $X_c$ denote $y, p_t$ and $\varphi$. In this
way, we obtain a flat distribution, i.e. $\rho(X_c)=1$.

\section{The results}

A 5000-event sample for central Au-Au collisions at $\sqrt
{s_{\rm{NN}}} =200$~GeV was produced by the AMPT model. The
corresponding results of a self-similar analysis in l-, 2- and
3-dimensional phase space are presented in Figs.1, 2 and 3. The
range for the three phase-space variables, rapidity $y$, azimuthal
angle $\varphi$ and transverse momentum $p_t$, were chosen as
($-6\leqslant y\leqslant6$, $0\leqslant pt \leqslant 3$ GeV, $0
\leqslant \varphi \leqslant 2\pi$).

\subsection{One-dimensional analysis and Hurst parameters}

Fig. 1 shows the results from a one-dimensional analysis of NFM vs.
M, which the partitioning $M = 1, 2, \cdots, 40$ was used for all
three variables ($M=M_{y}=M_{p_t}=M_{\varphi}$). Obviously, the
one-dimensional second-order NFM saturates for all the three cases
of $y, p_t$ and $\varphi$.

\begin{figure*}[htbp]
\begin{center}
\includegraphics[width=6cm,height=8cm]{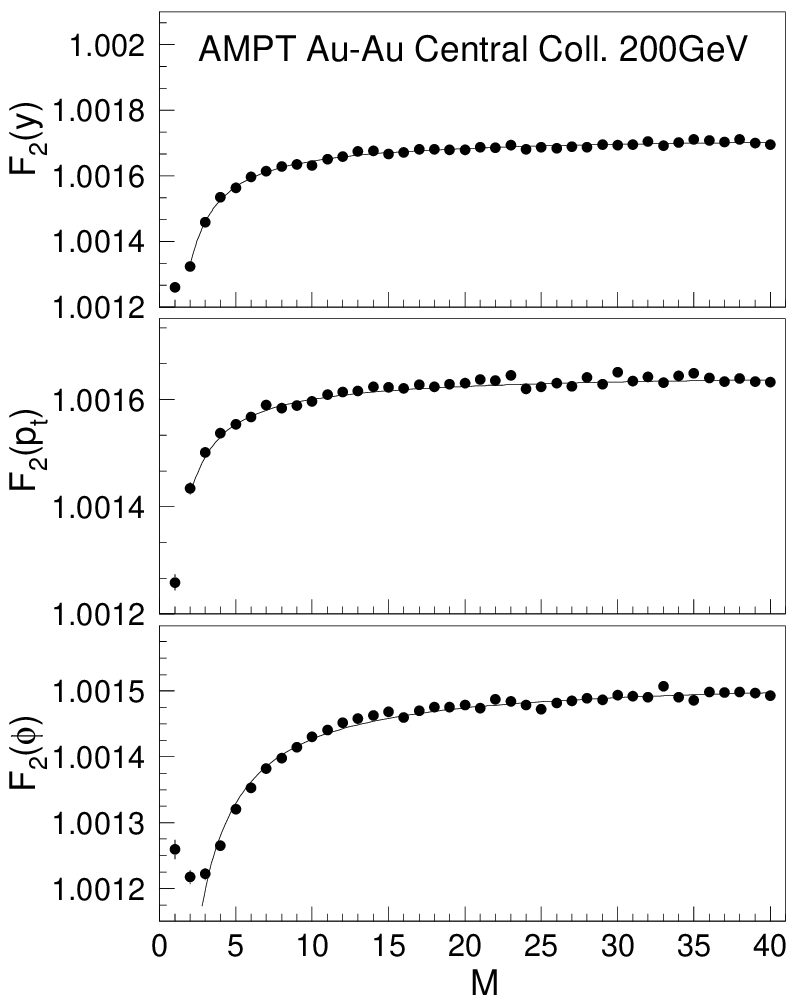}
\caption{Saturation curves for the second-order NFM of central Au-Au
collisions in the three one-dimensional variables indicated. The
curves are fitted by Eq. (6) after omission of the first point
(first two points in the case of $F_2(\varphi)$).}
\end{center}
\end{figure*}

It is easy to obtain the saturation exponent from Fig. 1, after
reducing the influence of momentum conservation~\cite{liu} by
excluding low values of M. The fitting results obtained according to
Eq. (6) are also shown in Fig. 1, considering three phase-space
variables. The corresponding parameter values are given in Table 1.

\begin{table*}[htbp]
\caption{Parameter values obtained from a fit by Eq. (6) for
one-dimensional NFM.}
\begin{center}
\begin{tabular}{ccccc}\hline
variable  & \multicolumn{1}{c}{A}& \multicolumn{1}{c}{B}&
\multicolumn{1}{c}{$\gamma$}&$\chi^2/DF$ \\\hline
$y$       &1.001720$\pm$0.000003 &0.0008$\pm$0.00003 &1.03$\pm$0.04 &39/37 \\
$p_t$     &1.001655$\pm$0.000004 &0.0004$\pm$0.00003 &0.91$\pm$0.07 &56/37 \\
$\varphi$ &1.001521$\pm$0.000004 &0.0010$\pm$0.00006 &1.03$\pm$0.05
&51/36 \\\hline
\end{tabular}
\end{center}
\end{table*}

As shown in Table 1, the saturation index $\gamma_y$ and
$\gamma_{\varphi}$ are equal, and they are slightly greater than
$\gamma_{p_t}$. The three exponents are approximately the same, i.e.
$\gamma_y = \gamma_{\varphi}\approx \gamma_{p_t} $, within the error
range. Accordingly, the Hurst exponents deduced from Eq. (7) are:
$$H_{yp_t}=0.94\pm0.05,\ \ H_{\varphi p_t}=0.94\pm0.06,\ \ H_{y\varphi}=1.00\pm0.04$$
From these Hurst exponents, we in fact obtain isotropic values
($H_{yp_t} = H_{\varphi p_t}\approx H_{y\varphi}$) for any two
directions of multiparticle production in three-dimensional phase
space. This means that fractality in multiparticle production of
central Au-Au collisions is self-similar.

\subsection{Two-dimensional analysis}

The plots for two-dimensional self-similar NFMs for orders $q=$ 2 -
9 in three different planes shown in Fig. 2, which have an isotropic
partition of the phase space, i.e. the Hurst exponent by $H_{p_ty}=
H_{p_t\varphi}=H_{y\varphi}=1$. We performed a linear fit to the
$\ln{F_q}$ vs. $\ln{M}$ using:
\begin{equation}
\ln{F_q}=c+\phi_q \ln{M},  
\end{equation}
which is derived from Eq. (2). The M $= M_{y}M_{\varphi}=
M_{\varphi}M_{p_t}= M_{y}M_{p_t}$ are the partitioning numbers of
two-dimensional phase space, and
$M_{y}=M_{\varphi}=M_{p_t}=1,2,\cdots,20$ are the partitioning
numbers in one dimension. The fitting curves are shown in Fig. 2 as
solid lines.

\begin{figure*}[htbp]
\begin{center}
\includegraphics[width=13cm]{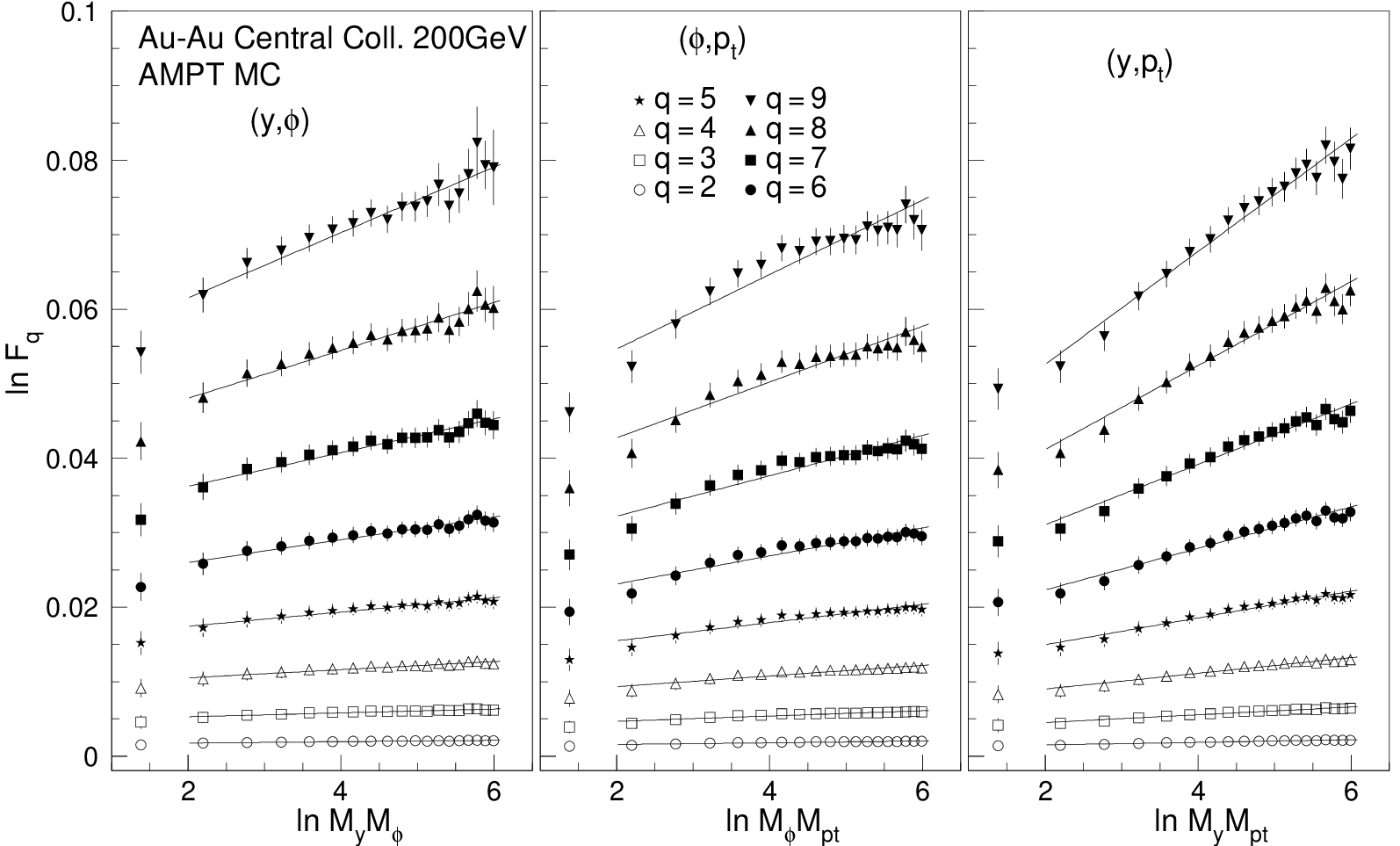}
\end{center}
\caption{Two-dimensional plots of $\ln{F_q}$ vs. $\ln{M}$ and the
results of the self-similar analysis with a linear fit in central
Au-Au collisions. The left-hand figure is for the ($y,\varphi$)
plane, the central figure is for the ($\varphi,p_t$) plane; the
right-hand figure is for the ($y,p_t$) plane.}
\end{figure*}

It can be seen that $\ln{F_q}$ increases linearly with increasing
$\ln{M}$, and the plots fit well with a straight line after the
first point is omitted in order to reduce the influence of momentum
conservation~\cite{liu}. Similar trends are found in the three
planes of the phase space for ($y,\varphi$),($\varphi,p_t$) and
($y,p_t$) in self-similar analysis.

In addition, $F_q(y,\varphi)$ presents a better scaling feature than
$F_q(\varphi,p_t)$ or $F_q(y,p_t)$, for the reason that the value of
Hurst exponents to partition phase space we take to be exactly equal
to 1 in all three phase-space planes, but the exponent equals 1
($H_{y\varphi}=1.00\pm0.01$) in the ($y,\varphi$) plane and is
approximately equal to 1 ($H_{yp_t}=0.94\pm0.05, H_{\varphi
p_t}=0.94\pm0.06$) in the ($\varphi,p_t$) and ($y,p_t$) planes.
Strictly according to the Hurst exponents in different directions,
if we partition the phase space using a non-integer
technique~\cite{chen4,chen1}, the scaling features would also be all
very precise.

\subsection{Three-dimensional analysis}

 We can perform a self-similar analysis in three-dimensional phase space, with the
Hurst exponents obtained above. For convenience, we approximate the
Hurst exponents by $H_{p_ty}= H_{p_t\varphi}=H_{y\varphi}=1$. From
Eq. (3), it follows that
$\lambda_y=\lambda_{p_t}=\lambda_{\varphi}$. We use a partitioning
$M=M_yM_{p_t}M_{\varphi}$, where $M_y=M_{p_t}=M_{\varphi} =
1,2,3,\cdots,12$.

The results for $\ln{F_q}$ vs. $\ln{M}$ in three-dimensional phase
space for orders $q=2 - 9$ are shown in Fig. 3. In order to show the
quality of the scaling law, linear fits according to Eq. (9) are
compared to the data in Fig. 3. The fitting results are listed in
Table 2. To reduce the influence of momentum conservation, the first
point are excluded in all the fits.

It can be seen from Fig. 3 that the results give a linear fit after
the first points for orders $q=2 -9$ are omitted. It is pointed out
that the fractral of final-state multiplicity production for the
central Au-Au collision at $\sqrt {s_{\rm{NN}}}=200$~GeV is
self-similar.

\begin{figure*}[htbp]
\begin{center}
\includegraphics[width=6.5cm]{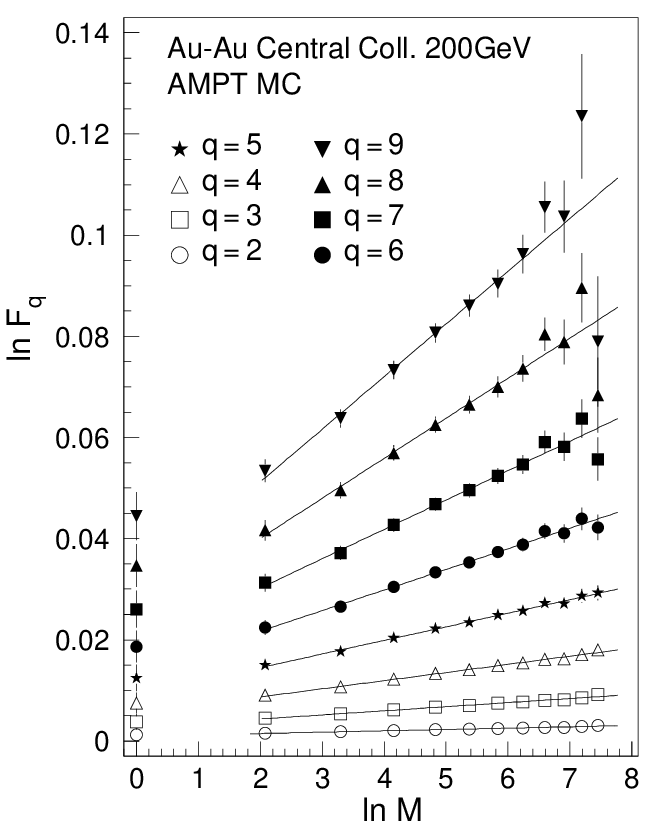}
\caption{Logarithmic distribution of three-dimensional NFM
self-similar analysis in central Au-Au collisions at $\sqrt
{s_{NN}}=200$~GeV for the order $q=2-9$. The curves are fitted by
Eq.(9)}
\end{center}
\end{figure*}

\begin{table*}[htbp]
\caption{Parameter values obtained from a fit of the
three-dimensional NFM by Eq. (9). }
\begin{center}
\begin{tabular}{cllc}\hline
q  & \multicolumn{1}{c}{$c$}&\multicolumn{1}{c}{$\phi_q$}&
\multicolumn{1}{c}{$\chi^2/DF$}\\\hline
2 &0.0009$\pm$0.0004&0.0003$\pm$0.0001   &0.2/8  \\
3 &0.0027$\pm$0.0006&0.0008$\pm$0.0001   &0.7/8  \\
4 &0.0055$\pm$0.0009&0.0016$\pm$0.0002   &0.8/8  \\
5 &0.0091$\pm$0.0012&0.0027$\pm$0.0002   &0.6/8  \\
6 &0.0136$\pm$0.0014&0.0041$\pm$0.0003   &2/8  \\
7 &0.0186$\pm$0.0019&0.0058$\pm$0.0004   &4/8  \\
8 &0.0242$\pm$0.0023&0.0079$\pm$0.0005   &8/8  \\
9 &0.0304$\pm$0.0028&0.0104$\pm$0.0006   &10/8  \\\hline
\end{tabular}
\end{center}
\end{table*}

However, we can also consider that an effective fluctuation strength
can be taken as a characteristic quantity for the strength of
dynamical fluctuations~\cite{FJHPLB}, defined as:
\begin{equation}
\alpha_{eff}=\sqrt{\frac{6\ln2}{q}(1-D_q)}=\sqrt{\frac{6\ln2}{q}\frac{\phi_q}{q-1}},
\end{equation}
where $q$ is the order of NFM, and $\phi_q$ is the intermittency
exponent. We calculate the effective fluctuation strengths
$\alpha_{eff}$ by Eq. (10) in central Au-Au collisions. The results
are listed in Table 3. For comparison, Table 3 also lists the
effective fluctuation strengths in NA22's hadron-hadron ($\pi^+
p+K^+p$) collision experiment at 250 GeV and L3's e$^+$e$^-$
collision experiment at 91.2 GeV, where the experimental
intermittency exponents for these effective fluctuation strengths
are taken from~\cite{chen1,NA22,Chen}.

In Table 3 it can be seen that, firstly, the effective fluctuation
strengths are constant when the intermittent exponents increase with
the order $q$ increasing; secondly, the effective fluctuation
strengths in heavy ion collisions are less than those in
hadron-hadron ($\pi^+ p+K^+p$) collisions and e$^+$e$^-$ collisions.
In other words, the dynamic fluctuation in relativistic heavy ion
collisions is much smaller than for hadron-hadron and e$^+$e$^-$
collisions.

\begin{table*}[htbp]
\begin{center}
\caption{Comparison of effective fluctuation strengths for
hadron-hadron, e$^+$e$^-$ and Au-Au collisions. }
\begin{tabular}{cccc}\hline
&   \multicolumn{3}{c}{$\alpha_{eff}$}\\\cline{2-4} q  &
\multicolumn{1}{c}{Au-Au}&\multicolumn{1}{c}{$\pi^+(K^+)p$
~\cite{chen1,NA22}}&
\multicolumn{1}{c}{$e^+e^-$~\cite{Chen}}\\\hline
2 & 0.025$\pm$0.004&0.356$\pm$0.011   &0.635$\pm$0.010   \\
3  & 0.024$\pm$0.001 &0.408$\pm$0.011   &0.644$\pm$0.013 \\
4 & 0.024$\pm$0.001&0.496$\pm$0.012   &0.612$\pm$0.009  \\
5  & 0.024$\pm$0.001&0.572$\pm$0.017   &0.600$\pm$0.008  \\
6&0.024$\pm$0.001& *& * \\
7&0.024$\pm$0.001& *& * \\
8&0.024$\pm$0.001& *& * \\
9&0.025$\pm$0.001& *& * \\\hline
\end{tabular}
\end{center}
\end{table*}

\section{The discussion of FM's scaling properties}

We can also consider the formula~\cite{Hwa1}
\begin{equation}
F_q \propto  F_2^\beta,\end{equation} and
\begin{equation}
 \beta_q \propto (q-1)^\nu
\end{equation} to check scaling property of our $F_q$
obtained in the AMPT model, where $\beta_q =\phi_q/\phi_2$. We draw
a plot of $\ln F_q$ vs $\ln F_2$ similar to Fig.~3 shown in the
Fig.~4. In order to show the quality of the scaling law, linear fits
according to Eq.~(11) are compared to the data in Fig. 4. It is
further shown that the system of final-state multiplicity production
for the central Au-Au collision at $\sqrt {s_{NN}} = 200$~GeV
exhibits good scaling properties, i.e its fractral is self-similar.
\begin{figure*}[htbp]
\begin{center}
\includegraphics[width=6.cm]{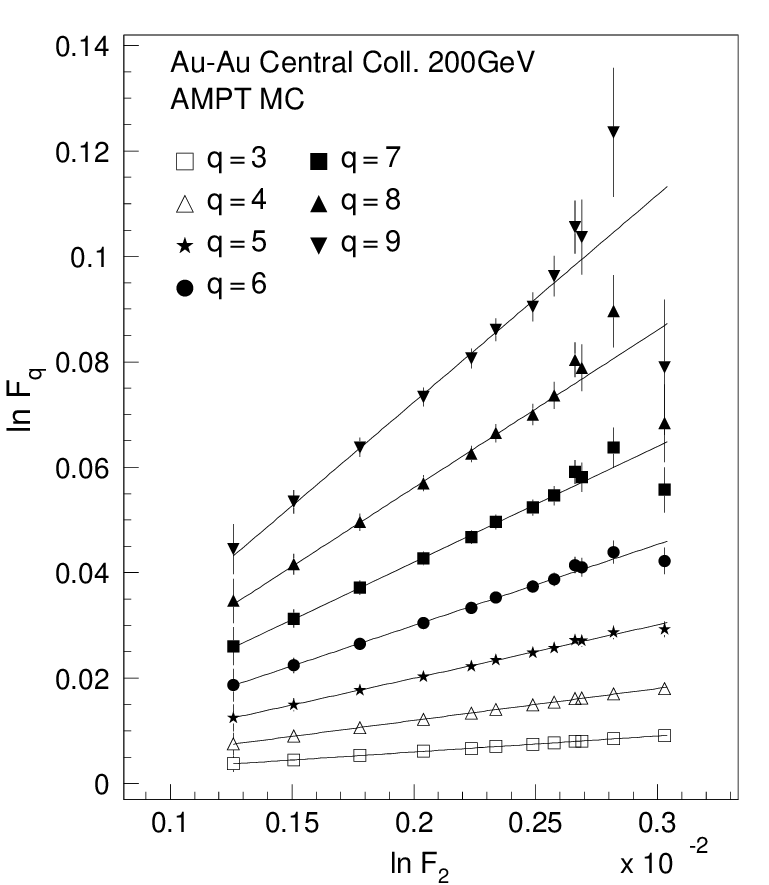}
\caption{The distribution of 3D $\ln F_q$ vs $\ln F_2$ in central
Au-Au collisions at $\sqrt {s_{NN}}=200$~GeV. The curves are fitted
by Eq.(11)}
\end{center}
\end{figure*}
The $\nu$ parameter may characterizes all the intermittency indices
derived in any particular analysis. If in certain region of phase
space where the hadronization of partons created in heavy-ion
collisions can be described by the Ginzburg-Landau type of phase
transition, then the factorial moments analyzed in that region can
be related as in Eq.(12), with $\nu = 1.304$~\cite{Hwa2}. Using the
data from Table 2, a plot of $\beta_q(\phi_q/\phi_2)$ vs $q$ show in
Fig.~5(solid circle points), with fitting parameter $\nu_{y\varphi
p_t}=1.86\pm 0.07$. We also compare the two-dimensional results with
the three-dimensional result using same way shown in Fig.~5(the
hollow symbols), with the fitting parameters $\nu_{y\varphi}=1.85\pm
0.13,\ \nu_{\varphi p_t}=1.94\pm 0.10,\ \nu_{yp_t}=1.94\pm 0.19$.
Obviously, the two-dimensional and three-dimensional results are
equal within the error range.

It is noteworthy that there is a big difference between our model
result and the value of Ginzburg-Landau type of phase transition.
The fact that $\nu_{y\varphi p_t}= 1.86\pm 0.07$ is larger than
1.304 indicates that the fluctuations simulated by AMPT are actually
larger than those due to the Ginzburg-Landau type of phase
transition, even though Table 3 suggests that those fluctuations are
weaker than the ones in hadronic and leptonic collisions. This is a
result worthy of our interest in further exploration. But it may be
because the AMPT model does not include the physics of phase
transition.

\begin{figure*}[htbp]
\begin{center}
\includegraphics[width=6.5cm]{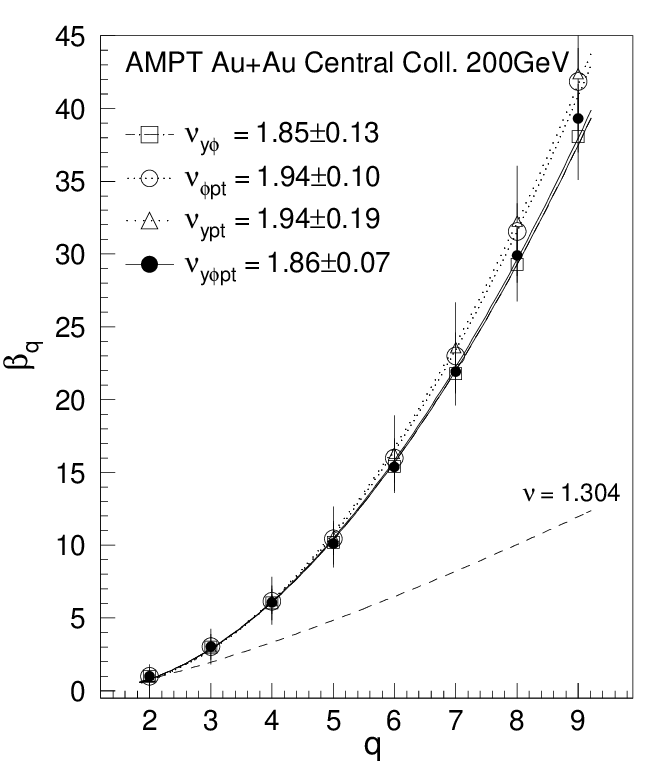}
\caption{Parameter $\beta_q$ vs $q$. The open symbols represent
two-dimensional results from Fig.2 and the solid symbols are
three-dimensional results from Fig.3. The curves are fitted by
Eq.(12). The dashed line corresponds to the case when the
Ginzburg-Landau type of phase transition applies.}
\end{center}
\end{figure*}

It is known that hadronization of partons may occur at different
times in the evolution of the system and may populate different
$p_t$ intervals, depending on the time of hadronization. If we make
factorial moment analysis by splitting the $p_t$ range to smaller
intervals of $0.5 \leqslant p_t < 1.,\ 1. \leqslant p_t < 1.5,\ 1.5
\leqslant p_t < 2.5,\ 2.5 \leqslant p_t < 3.5,\ 3.5 \leqslant p_t <
4.5$~GeV/C, this different $p_t$ slices may avoid the overlapping of
multiplicities of hadronization products in an given event, and
yield different $\nu$ values. The value of $\nu$ for the $p_t$
interval $1 \leqslant p_t < 1.5$~GeV/C is expected to very different
from the value in the interval $3.5 \leqslant p_t < 4.5$~GeV/C
because the latter is dominated by jets effects.

The results of two-dimensional factorial moment loglog distributions
in $(y, \varphi)$ plane by splitting the $p_t$ range to smaller
intervals as given above are shown in Fig.6. One can see that the
intermittency or fluctuations obviously increase gradually with the
increasing of transverse momentum $p_t$ from Fig.6(a) to (b), and
then (c), until (e). Nevertheless, the event multiplicity of Au-Au
collisions at $\sqrt {s_{NN}}=200$~GeV is not high enough to give a
well-determined value of $\nu$ for each $p_t$ interval. So the
quantitative analysis of factorial moment in Fig.6 can not be made
because of the limited statistics of particle numbers per event in
the $p_t > 2.5$~GeV/C, as shown in Fig.6(d) and (e).

\begin{figure*}[htbp]
\begin{center}
\includegraphics[width=12.5cm]{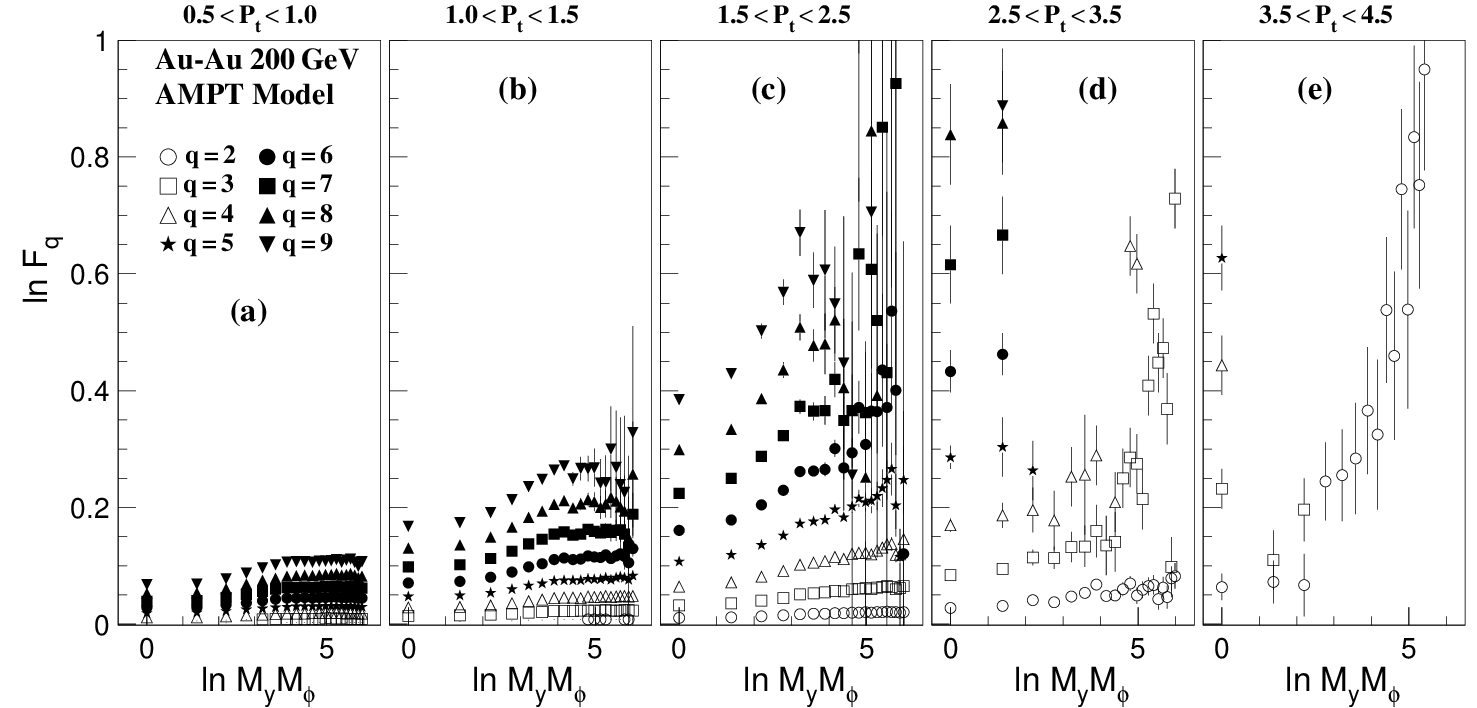}
\caption{The loglog distribution of two-dimensional factorial
moments $F_q$ in the $(y,\varphi)$ plane at various $p_t$
intervals.}
\end{center}
\end{figure*}

However, we can consider the factorial moment in the case of phase
space partition number $M=1$ as formula
\begin{eqnarray} f_q(y, \varphi)=
\frac{<n_m(n_m-1)\cdots(n_m-q+1)>}{<n_m>^q},
\end{eqnarray}
to analyze fluctuational property of final state multiparticle
system in Au-Au collisions at $\sqrt {s_{NN}}=200$~GeV. The result
of two-dimensional factorial moment in $(y, \varphi)$ plane by
splitting the $p_t$ range to smaller intervals calculated by Eq.(13)
are shown in Fig.7. Obviously, we can see from Fig.7 that the
factorial moment $f_q(y,\varphi)$ increases rapidly with the
increasing of transverse momentum $p_t$ indicating that the value of
fluctuation of final state multiparticle system in Au-Au collisions
for the $p_t$ interval $3.5 \leqslant p_t < 4.5$~GeV/C is much
larger than the value in the interval $1. \leqslant p_t <
1.5$~GeV/C. These results are consistent with our expectations.

\begin{figure*}[htbp]
\begin{center}
\includegraphics[width=5.5cm]{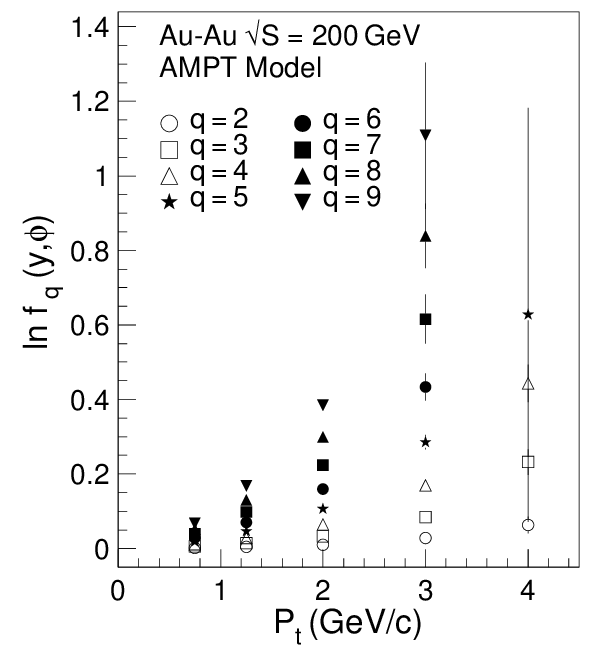}
\caption{The distribution of two-dimensional factorial moment $\ln
f_q(y,\varphi)$, as a function of $p_t$, with partition number
$M=1$.}
\end{center}
\end{figure*}
\section{Summary}
This paper describes a self-similar analysis of the factorial moment
for the order $q=2-9$ in three-dimensional phase space using the
default AMPT model to generate 5000 central Au-Au collisions at
$\sqrt {s_{\rm{NN}}}=200$~GeV. By one-dimensional projection of the
factorial moments with the Ochs saturation formula Eq. (6), the
Hurst exponents were derived for all combinations of the phase-space
variables as ($y$, $p_t$, $\varphi$), which are almost identical and
approximately equal to 1, i.e. $H_{p_ty}= H_{p_t\varphi}\simeq
H_{y\varphi}=1$. Therefore, we conclude that fractality in
multi-particle production of central Au-Au collisions is indeed
self-similar. Furthermore, the three-dimensional self-similar
analysis shows good scaling behavior.

Finally, the FM's scaling property and the Ginzburg-Landau type of
phase transition applies are discussed. We measured the parameter
$\nu$ which characterizes the intermittency indices derived in
particular analysis. It is found that there is a big difference
between our model result $\nu_{y\varphi p_t}= 1.86\pm 0.07$ and the
value of Ginzburg-Landau type of phase transition $\nu=1.304$. The
fact that our model result is larger than 1.304 indicates that the
fluctuations simulated by AMPT are actually larger than those due to
the Ginzburg-Landau type of phase transition. This is worthy of our
interest in further exploration. It may be because the AMPT model
does not include the physics of phase transition. We also explored
the intermittency and fluctuation in dependence on the transverse
momentum. The result shows that the factorial moment, as well as
intermittency or fluctuations, increases rapidly with the increasing
of transverse momentum $p_t$.

It should be noted that our results obtained should be checked by
corresponding analysis of the experimental data. If they turn out to
disagree, that would indicate the need to modify AMPT, which has
been tuned to agree with most features of the data, but not the
fluctuations in bin and event multiplicities.

\begin{center} {ACKNOWLEDGEMENT} \end{center}
This work is supported by Fundametal Research Funds for NSFC
(11305144, 11303023) and Central Universities (GUGL
100237,120829,130249) in China. The authors thank Professor Wu
Yuan-fang and Yang Chun-Bin in the Particle Physics Research
Institute of Huazhong Normal University for helpful discussions.

\end{document}